\documentstyle[twocolumn,prc,aps,psfig]{revtex}
     \newcommand{\pathnow}{}

\begin{document}\hbadness=10000
\twocolumn[\hsize\textwidth\columnwidth\hsize\csname %
@twocolumnfalse\endcsname
\title{Quark-Gluon Plasma Fireball}
\author{Salah Hamieh, Jean Letessier and Johann Rafelski}
\address{
Department of Physics, University of Arizona, Tucson, AZ 85721\\
{\it and}
Laboratoire de Physique Th\'eorique et Hautes Energies,
Universit\'e Paris 7, 2 place Jussieu, F--75251 Cedex 05.
}
\date{Received: June 17, 2000}
\maketitle
\begin{abstract}
\noindent Lattice-QCD results provide an opportunity to model, and extrapolate
to finite baryon density, the properties of the quark-gluon plasma (QGP). 
 Upon fixing the scale of the thermal coupling
constant and vacuum energy to the lattice data, the properties of resulting 
QGP equations of state (EoS) are developed. We show that the physical properties 
of the dense matter fireball formed in heavy ion collision experiments at CERN-SPS
are well described by the QGP-EoS we presented. We also
estimate the properties of the fireball  formed in early stages of 
nuclear collision, and argue that QGP formation must be expected 
down to 40$A$ GeV in central Pb--Pb interactions.\\

\end{abstract}
\vspace{-0.35cm}
]
\begin{narrowtext}
\section{Introduction}\label{intro}
It is believed today that a new state of matter has been formed 
in relativistic nuclear collisions at CERN \cite{CERN}.
The existence of a novel non-nuclear high temperature phase of 
elementary hadron matter, consisting of deconfined quarks and gluons, 
arises from  the current knowledge about quantum chromodynamics (QCD),
the theory of strong interaction. At sufficiently high temperature, the 
strong interactions weaken (asymptotic freedom), and thus
we expect that hot and dense nuclear matter will behave akin to a free gas 
of quarks and gluons \cite{Col75}.  At issue is today if the new phase of
matter observed at CERN is indeed this so called  quark-gluon plasma
(QGP) phase. One of the ways to test, and possibly falsify, the QGP hypothesis
is to consider if the expected properties of the QGP indeed agree 
with experimental data which have been at the center 
of the CERN announcement. 

The observational output of experiments we consider
are particle abundances and particle spectra.
We address here results on hadron and strange hadron production  
at the equivalent center of momentum (CM) reaction energy 
$E_{\rm CM}=8.6$\,GeV per baryon in Pb--Pb reactions.  
As seen in many experimental results \cite{CERN}, which we will
not restate here in further detail, in these high energy 
nuclear collisions a localized dense and hot matter fireball is formed.
In our earlier analysis of experimental results \cite{Let00,Let99}, 
we have obtained diverse physical properties of the
source, such as energy per baryon content $E/b$, 
entropy per baryon content $S/b$, and last, not least, 
strangeness content per baryon $\bar s/b$. These values are
associated with temperature $T_f$ and 
baryo-chemical potential $\mu_b$ at  which these hadron abundances 
are produced (chemical freeze out), along with other 
chemical properties and the  collective velocity 
of the matter emitting these particles $v_c$.

Our main aim is thus to compare the 
properties of the QGP phase, modeled to agree with 
the lattice-QCD calculations \cite{Kar00}, to the 
properties of the fireball obtained in the study of 
hadronic particle abundances. We thus develop
 semi-phenomenological QGP equations of state (EoS)
based on thermal and lattice-QCD results. 
These are then applied to explore, in a systematic fashion, properties of the
fireball formed in the Pb-Pb reactions.  
We show that the physical properties of the fireball extracted from particle 
production data are the same as obtained employing  our QGP-EoS.  

Using our QGP-EoS,  we also explore  the initial thermal conditions 
reached in the collision both in, and out, of quark chemical equilibrium. 
 We have already shown previously
that the strangeness yield is  following the predicted QGP yield \cite{Let00,Let99}. 
We also show that, when the 
energy and baryon number deposited in the initial fireball drops below 
20\%, {\it e.g.\/}, due to large impact parameter 
interactions, or/and  small nuclei involved 
in the collision, the formation of the QGP phase becomes improbable.

An important dynamical  aspect of our experimental data 
analysis \cite{Let00,Let99}, on which this work relies, is the
sudden disintegration (hadronization) of the 
 fireball into the final state hadrons. This reaction mechanism is 
a priori not very surprising, since a fireball formed in these collisions 
explodes,  driven by internal compression pressure. However, we did not 
find in the particle production analysis as the particle source the expected 
chemically equilibrated,  confined hadron phase. Furthermore, the particle 
production temperature (chemical freeze-out) we found corresponds to a deeply
super-cooled state, which can be subject to mechanical instability \cite{Cso94}. 
We have checked elsewhere this assertion, applying the EoS developed 
here in a detailed  analysis of the fireball  sudden break-up, and have found 
that it occurs at the condition of mechanical instability~\cite{Raf00}. 

A consequence of the sudden fireball break up is that akin to the situation 
found in the study of properties of the early universe,
particle abundances do not reach chemical equilibrium. 
Our analysis accounts in full for this important fact \cite{Let00,Let98}. 
For an appropriate criticism of the  chemical 
equilibrium models, and a list of related work, 
we refer the reader to work of  Bir\'o \cite{Bir00}. 

In the next section, we address the thermal 
QCD interaction coupling $\alpha_s$ we will use. 
In section \ref{SecQGP}, we define the equations of state and
the parameters of our approach and explore  which 
is the best scheme for the extrapolation of the 
lattice data. In section \ref{QGPprop}, we 
present properties of the QGP phase, relevant both 
to the study of the freeze-out conditions, and the study of the 
initial conditions reached in the collision.  We present and discuss the 
comparison of the properties of the exploding fireball with those
measured by means of hadron production in section \ref{analyze}.
Our conclusions follow in 
section \ref{conclusion}. 

\section{QCD Interactions in plasma}\label{QCDIntr}
The energy domain in which we explore diverse properties of dense strongly 
interacting matter is  barely above the scale 1\,GeV and thus, in  our
consideration, an important input is the scale $\mu$ dependence of the 
QCD coupling constant, $\alpha_s(\mu)$, which we obtain solving
\begin{equation}
\label{dalfa2loop}
\mu \frac{\partial \alpha_s}{\partial \mu}=
-b_0\alpha_s^2-b_1\alpha_s^3+\ldots \equiv \beta^{\mbox{\scriptsize pert}}_2\,.
\end{equation}
$\beta^{\mbox{\scriptsize pert}}_2$ is
the beta-function of the renormalization group 
in two loop approximation, and 
$$b_0=\frac{11-2n_{\mathrm f}/3}{2\pi}\,,\quad 
   b_1=\frac{51-19n_{\mathrm f}/3}{4\pi^2}\,.$$ 
$\beta^{\mbox{\scriptsize pert}}_2$
does not depend on the renormalization scheme,
and solutions of Eq.\,(\ref{dalfa2loop}) differ from higher 
order renormalization scheme
dependent results  by less than the error introduced by the experimental 
uncertainty in the measured value of $\alpha_s(\mu=M_Z)=0.118+0.001-0.0016$. 
When solving  Eq.\,(\ref{dalfa2loop}) with this initial condition, 
we cross several flavor mass thresholds and thus 
$n_{\mathrm f}(\mu)$ is not a constant in the interval $\mu\in(1,100)$\,GeV.
Any error made when not properly accounting for  
$n_{\mathrm f}$  dependence on  $\mu$
accumulates in the solution of  Eq.\,(\ref{dalfa2loop}). In consequence,
a popular approximate analytic solution shown as function of $T$,
as dotted line in figure \ref{THERMalpha}, 
\begin{equation}
\label{Lambdarun}
\alpha_s(\mu)\simeq{2 \over b_0 \bar{L}}\left[1-
\frac{2 b_1}{b_0^2}\frac{\ln\bar{L}}{\bar{L}}\right],\quad
\bar{L}\equiv\ln(\mu^2/\Lambda^2)\,,
\end{equation}
with $ \Lambda= 0.15$\,GeV and $n_f=3$, 
is not precise enough  compared to the exact 2-loop numerical solution.

We show the numerical solution for $\alpha_s(\mu)$ 
in figure \ref{THERMalpha},  setting 
\begin{eqnarray}\nonumber
\mu=2\pi T=\kappa {T}/{T_c},\quad \kappa=1\mbox{\,GeV}\,,
\end{eqnarray}
where the solid line corresponds to 
$\alpha_s(\mu=M_Z)=0.118$, bounded by the experimental 
uncertainty \cite{LTR96}. 
We observe that  $\alpha_s/\pi<0.25$ is relatively small. 
However, the  expansion parameter of thermal 
QCD  $g=\sqrt{4\pi \alpha_s}>1$\,. Therefore 
even the small difference shown in figure \ref{THERMalpha}
between the approximation (dotted line) used in above references 
and the  exact  result is  relevant.    A large value 
of $g$ has been a source of considerable concern about 
validity of perturbative expansion in 
thermal QCD \cite{Arn94,Zha95,Bra96,And99}. 
 
\begin{figure}[tb]
\vspace*{-2.5cm}
\hspace*{-1.cm}\psfig{width=11.4cm,clip=,figure=\pathnow 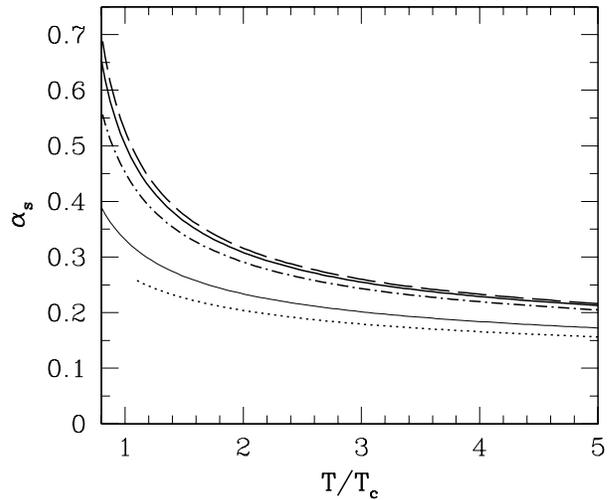}
\vspace*{-2.6cm}
\caption{ 
$\alpha_s(2\pi T)$ for $T_c=0.16$\,GeV. 
Dashed line: $\alpha_s(M_Z)=0.119$; solid line $=0.118$;
dot-dashed line $=0.1156$.
Dotted line: approximate 2-loop solution given in Eq.(\protect\ref{Lambdarun}).
Thin solid line: same as dotted, but extrapolating 
with $n_f=5$ to $\alpha_s(M_z)=0.118$\,.
\label{THERMalpha}} 
\end{figure}

The analytical function,
 Eq.\,(\ref{Lambdarun}), is in fact a very good solution of Eq.\,(\ref{dalfa2loop}).
For $\Lambda=0.225$\,GeV and $n_f=5$, it agrees with numerical result for the
entire range, $M_Z=92$\,GeV$>\mu> M_B\simeq 4.5$\,GeV, in which $n_f=$ Const., here
$M_B$ is the bottom quark mass. 
However, the region of interest for the QGP equations of state we explore is 
1\,GeV$<\mu<M_B$, where $n_f$ varies and the approximation,  Eq.\,(\ref{Lambdarun})
is not adequate.   The thin solid line, in  figure \ref{THERMalpha},
shows the behavior of the analytical solution with $n_f=5$ kept constant,
and for $\Lambda=0.225$\,GeV which assures the boundary value 
$\alpha_s(M_Z)=0.118$.

\section{The quark-gluon liquid}\label{SecQGP}
We now can define the  quark-gluon liquid model which
describes well the properties of QGP
determined by lattice-QCD method. \\
{\bf 1.}
To relate the QCD scale to the temperature $T=1/\beta$, 
 we use for the scale the Matsubara frequency \cite{Pes00}:
\begin{equation}
\label{runalTmu}
\mu=2\pi \beta^{-1}\sqrt{1+\frac{1}{\pi^2}\ln^2\lambda_{\mathrm q}}
=2\sqrt{(\pi T)^2+\mu_{\mathrm q}^2}\,.
\end{equation}
This extension to finite chemical  potential $\mu_{\mathrm q}$, or 
equivalently quark fugacity  $\lambda_{\mathrm q}=\exp{\mu_{\mathrm q}/T}$, 
is motivated by the form of plasma frequency  entering
the computation of the vacuum polarization function \cite{Vij95}.
In principle, there should be in Eq.\,(\ref{runalTmu}) also a contribution from the 
strange quark chemical fugacity $\lambda_{\mathrm s}$ expressed equivalently 
by the strange quark chemical potential  $\mu_{\mathrm s}$. However since 
strangeness conservation virtually assures that $\mu_s\simeq 0$, or equivalently, 
$\lambda_{\mathrm s}\simeq 1$, we will not pursue this further.\\
{\bf 2.}
To reproduce the lattice results 
available at $\mu_q=0$ \cite{Kar00},  we need
to introduce, in the domain of freely mobile quarks 
and gluons, a finite vacuum energy  density:
$${\cal B}=0.19\,\frac{\mbox{GeV}}{\mbox{fm}^3}\,.$$
This also implies, by virtue of relativistic invariance,
that there must be a (negative) 
associated pressure acting on the surface of this volume, 
aiming to reduce the size of the deconfined region.  
These two properties of the vacuum follow
consistently from the vacuum partition function:
\begin{equation}
\label{Zbag}
\ln{\cal Z}_{\mbox{\scriptsize vac}}\equiv -{\cal B}V\beta\,.
\end{equation}
{\bf 3.}
The partition function of the quark-gluon liquid comprises interacting 
gluons, $n_{\mathrm q}$ flavors of light quarks  \cite{Chi78}, 
and the vacuum ${\cal B}$-term. We
incorporate further the strange quarks by assuming that their mass 
in effect reduces their effective number  $n_{\mathrm s}<1$:
\begin{eqnarray}
\label{ZQGPL}
&&\frac{T}{V}\ln{\cal Z}_{\mathrm QGP}
\equiv P_{\mathrm QGP}=
-{\cal B}+\frac{8}{45\pi^2}c_1(\pi T)^4  \nonumber\\ 
&&+
\frac{n_{\mathrm q}}{15\pi^2}
\left[\frac{7}{4}c_2(\pi T)^4+\frac{15}{2}c_3\left(
\mu_{\mathrm q}^2(\pi T)^2 + \frac{1}{2}\mu_{\mathrm q}^4
\right)\right]
\nonumber\\ 
&&+
\frac{n_{\mathrm s}}{15\pi^2}
\left[\frac{7}{4}c_2(\pi T)^4+\frac{15}{2}c_3\left(
\mu_{\mathrm s}^2(\pi T)^2 + \frac{1}{2}\mu_{\mathrm s}^4
\right)\right]\,,
\end{eqnarray}
 where:
\begin{eqnarray}
\label{ICZQGP}
c_1&=&1-\frac{15\alpha_s}{4\pi}+ \cdots\,,\\ \nonumber
c_2&=&1-\frac{50\alpha_s}{21\pi}+ \cdots\,,\qquad
c_3=1-\frac{2\alpha_s}{\pi}+ \cdots\,.
\end{eqnarray}

In figure \ref{figPalfa1}, the `experimental' values are from 
numerical lattice simulations of  $P/T^4$ \cite{Kar00}.
 For practical reasons the lattice results for
`massless' 2 and 3-flavors were obtained with $m/T=0.4$, which 
reduces the particle numbers by 2\%, and this effect is allowed 
for in the quark-gluon liquid lines (two flavors: dashed, three 
flavors: dotted) in figure \ref{figPalfa1}.
In the case 2+1 flavors, a renormalized 
strange quark mass $m_s/T=1.7$, the 
lattice input has been $m_s^0/T=1$. 
This leads to
a $\simeq 50$\% reduction in strange quark number,
thus $n_{\mathrm s}\simeq 0.5$ will be used in 
our study of QGP properties, assuming that strangeness
is fully chemically equilibrated, unless otherwise noted. 
Thus, in general, we have 
$n_{\mathrm f}=n_{\mathrm q}+n_{\mathrm s}\simeq 2.5$.

The thin lines, in figure \ref{figPalfa1}, 
correspond to the previously reported results for a quark-gluon
gas \cite{And99},  with first order QCD correction introduced using 
approximate value of $\alpha_s(\mu)$, Eq.\,(\ref{Lambdarun}), 
without the vacuum pressure term. For $T\ge 2T_c$, 
the disagreement is an artifact of 
the approximation for $\alpha_s$.

Lattice results were also informally reported \cite{CERN} for the
energy density, with $n_f=3$, and are show in figure \ref{figendens25},
upper `experimental' points. The energy density,
\begin{equation}
\label{enerden0}
\epsilon_{\mbox{\scriptsize QGP}}=-
\frac{\partial\ln {\cal Z}_{\mathrm QGP}(\beta,\lambda)}{V\partial \beta}\,,
\end{equation}
is sensitive to the slope of the partition function. We show, in  
figure \ref{figendens25}, the energy density and pressure for two
extreme theoretical approaches:   the 
solid lines are  for the model we described above ($\mu=2\pi T,\,
 {\cal B}=0.19$\,GeV/fm$^3$),  the dotted lines are 
obtained in the thermal expansion, including all, up to fifth order,  
scale dependent terms obtained by Zhai and Kastening \cite{Zha95}, 
and choosing the scale $\mu=2.6\pi T$ 
in free energy and in the coupling constant, 
so the result  reproduces the pressure well (bottom dotted line in 
figure \ref{figendens25}). 

\begin{figure}[tb]
\vspace*{-2.9cm}
\hspace*{-1.4cm}\psfig{width=12.2cm,clip=,figure=\pathnow 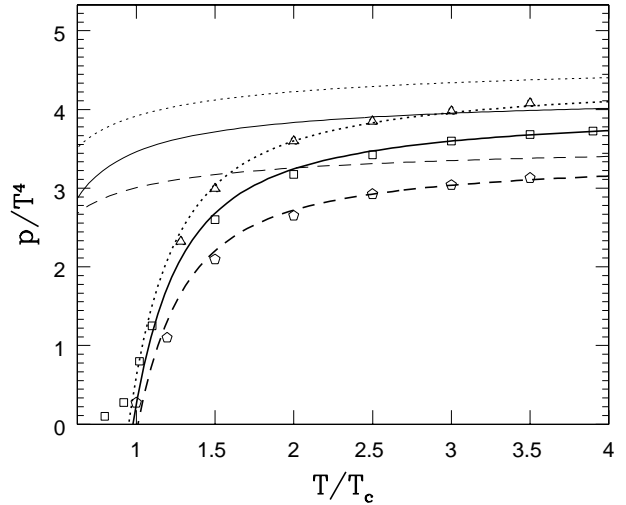}
\vspace*{-2.9cm}
\caption{ 
Lattice-QCD results \protect\cite{Kar00} for $P/T^4$ at $\lambda_q=1$,
compared  with our quark-gluon liquid model (thick lines) and lowest order
perturbative QCD using approximate $\alpha_s$, 
Eq.\,(\protect\ref{Lambdarun}) (thin lines 
used in prior work \protect\cite{Arn94,Zha95,Bra96,And99},): 
dotted line 3 flavors, solid line 2+1 flavors, 
and dashed line 2 flavors.\label{figPalfa1}
}
\end{figure}

In the upper portion of figure \ref{figendens25}, comparing these 
two  theoretical  approaches chosen to reproduce the pressure, 
we see  a  clear difference  in the energy density. The 5th 
order energy density (dotted line) 
$(g/4\pi)^5=(\alpha_s/\pi)^{5/2}/32$ \cite{Zha95},
disagrees with the lattice data also at high $T$, where these are
most precise. Near to $T\simeq 1.5T_c$, the solid line is visibly
better describing the pressure.  In fact, systematic study of the 
behavior of the $(g/4\pi)^n$ expansion reveals that higher order terms 
do not lead to a stable
result for the range of temperatures of interest to us \cite{Zha95,And99}.

It is not uncommon to encounter in a perturbative
expansion a semi-convergent series. The issue then is how to 
establish a workable scheme. Our result implies  that the choice of the 
Matsubara frequency $2\pi T$ as the scale $\mu$ 
of the running coupling constant has 
for yet unknown reasons, the effect
to minimize the contribution of the sum of all 
higher order terms in the expansion Eq.\,(\ref{ICZQGP}),  
even for moderate temperatures, 
once the vacuum pressure is introduced. 
The agreement between our model and the lattice 
calculations arises despite (or maybe because) the fact that in 
our evaluation of the temperature dependence of the 
coupling constant $\alpha_s(T)$  we ignored the dependence of 
$\beta^{\mbox{\scriptsize pert}}_2$ 
on the ambient temperature scale \cite{Elm95}. 
It seems that in fact the omission of this dependence  
conspires with omission of the higher order thermal contributions 
in the partition function, yielding the remarkable agreement 
we see between the lattice results and our simple minded approach,
which considers the first order corrections in $\alpha_s$ only.

\begin{figure}[tb]
\vspace*{-2.5cm}
\hspace*{-1.3cm}\psfig{width=11.8cm,clip=,figure=\pathnow 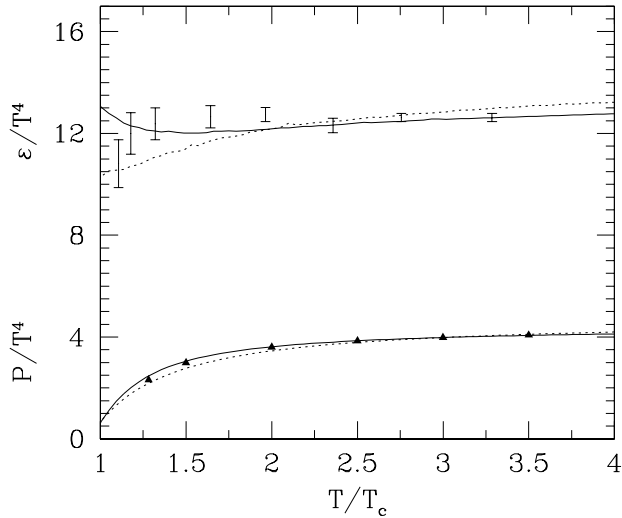}
\vspace*{-2.8cm}
\caption{ 
Top: informal lattice-QCD results \protect\cite{CERN} 
for $\varepsilon/T^4$ at $\lambda_q=1$,
compared  with our quark-gluon liquid model (solid line). 
Doted line is an alternative approach in which all terms in 
partition function are summed as given in \protect\cite{Zha95} 
and the scale is set at $\mu=2.6\pi T$.
Bottom: published lattice pressure results \protect\cite{Kar00}
compared to the two approaches. Solid line: first order with 
the bag constant; dotted line: 5th order with $\mu=2.6\pi T$. 
\label{figendens25}
}
\end{figure}

Given that our model  adequately describes
both pressure and energy density leads us to believe that we also
have  an appropriate  extrapolation scheme for the QGP properties to 
finite baryon density -- but we could 
not test if the coefficient $c_3$, Eq.\,(\ref{ICZQGP}),
describes  well the behavior of the partition function for 
finite chemical potentials, as such lattice-QCD results 
are not available. 

In absolute terms, the model we now adopt for
further study reproduces the lattice results well, at the level 
of a few percent. It can only be hoped that the lattice results 
have reached that level of precision. In our approach, the value of $\cal B$ 
we obtain and employ is entirely dependent on the quality of the lattice 
results. However, in another work  \cite{Raf00},
we have  considered the dynamics of the expanding fireball 
and obtained an estimate for the value of $\cal B$ as defined here.
We found that ${\cal B}\ge 0.17\,\mbox{GeV/fm}^3$, consistent with the
value we employ here  ${\cal B}=0.19\,\mbox{GeV/fm}^3$.

\section{Properties of QGP-liquid}\label{QGPprop}
We are now ready to explore the physical properties of the quark-gluon liquid. 
The energy density is obtained from Eq.\,(\ref{ZQGPL}), recalling that the 
scale of the interaction is given by Eq.\,(\ref{runalTmu}):
\begin{eqnarray}
\label{enerden1}
&&  
\epsilon_{\mbox{\scriptsize QGP}}=
  4{\cal B}+3P_{\mathrm QGP}+A_{\mathrm g}+A_{\mathrm q}+A_{\mathrm s}\,, \\ 
A_{\mathrm g}\!&=&(b_0\alpha_s^2+b_1\alpha_s^3)\frac{2\pi}{3} T^4 \,,\\
A_{\mathrm q}\!&=& (b_0\alpha_s^2\!+b_1\alpha_s^3)
  \!\left[\frac{5\pi n_{\mathrm q}}{18} T^4\!+ \frac{n_{\mathrm q}}{\pi}
  \!\left(\mu_{\mathrm q}^2 T^2\! + \frac{\mu_{\mathrm q}^4}{2\pi^2}\right)\!\right],\\
A_{\mathrm s}\!&=& (b_0\alpha_s^2+b_1\alpha_s^3)
   \!\left[\frac{5\pi n_{\mathrm s}}{18} T^4+\frac{n_{\mathrm s}}{\pi}
   \!\left(\mu_{\mathrm s}^2 T^2 + \frac{\mu_{\mathrm s}^4}{2\pi^2}\right)\!\right].
\end{eqnarray}
A convenient way to obtain entropy and baryon density uses the 
thermodynamic potential ${\cal F}$:
\begin{equation}
\label{FQGP}
\frac{{\cal F}(T,\mu_q,V)}{V}=
-\frac{T}{V}\ln {{\cal Z}(\beta,\lambda_q,V)}_{\mathrm QGP}
=-P_{\mathrm  QGP} \,.
\end{equation}
The entropy density is:
\begin{eqnarray}
s_{\mathrm QGP}&=&-\frac{d{\cal F}}{VdT}\,,\\
&=&\frac{(n_{\mathrm q}+n_{\mathrm s})7\pi^2}{15}c_2 T^3+
 n_{\mathrm q}c_3\mu_{\mathrm q}^2 T
+n_{\mathrm s}c_3\mu_{\mathrm s}^2 T\nonumber \\
&+&\frac{32\pi^2}{45}c_1 T^3 +
(A_{\mathrm g}\!+A_{\mathrm q}\!+A_{\mathrm s})
 \frac{\pi^2 T}{\pi^2 T^2+\mu_{\mathrm q}^2}\,.
\label{sidens}
\end{eqnarray}
 Noting that baryon density is 1/3  of quark density,  
we have:
\begin{eqnarray}
\rho_{\mathrm b}
&=&
-\frac{1}{3}\frac{d{\cal F}}{Vd\mu_{\rm q}}\\
&=& \label{rhodens}
\frac{n_{\mathrm q}}{3}c_3\left\{
\mu_{\rm q} T^2 + \frac{1}{\pi^2}\mu_{\rm q}^3\right\} 
+\frac{n_{\mathrm s}}{3}c_3\left\{
\mu_{\rm s} T^2 + \frac{1}{\pi^2}\mu_{\rm s}^3\right\} 
\nonumber\\ 
&&\hspace*{0.5cm}+
\frac13 (A_{\mathrm g}+A_{\mathrm q}+A_{\mathrm s})
\frac{\mu_{\mathrm q}}{\pi^2 T^2+\mu_{\mathrm q}^2}\,.
\end{eqnarray}

We show properties of the quark-gluon liquid in a wider range of parameters in 
figure \ref{EOSSfix}. We study the properties 
at fixed entropy per baryon $S/b$ since
an isolated ideal particle fireball would evolve at a fixed $S/b$. 
We consider the range $S/b=10$ (at left for the top panel,
 baryo-chemical potential $\mu_b$, and middle panel
baryon density $n/n_0$, here  $n_0=0.16/\mbox{fm}^3$, and bottom left for 
the energy per baryon $E/b$) in step of 5 units, up to maximum of $S/b=60$. 
The  highlighted curve, in figure \ref{EOSSfix},
is for the  value $S/b=42.5$, which value follows from earlier study of
hadronic particle spectra \cite{Let00}. 
The dotted line, at the 
minimum of $E/b\vert_{S/b}$, is where the vacuum and quark-gluon gas pressure balance. 
This is the equilibrium point and indeed  the energy per 
baryon does have a relative minimum there. 

Unlike the case for an ideal
quark-gluon gas, the lines of fixed $S/b$, seen in the
top panel of figure \ref{EOSSfix}
are not corresponding to $\mu_b/T=$\,Const., 
though for large $T$ and small $\mu_b$ they do show 
this asymptotic behavior. 
Since little entropy is produced during the evolution of 
the QGP fireball, the thick line in the lower panel of figure \ref{EOSSfix}
describes the approximate trajectory in time of the fireball
made in Pb--Pb interactions at the projectile energy 158$A$ GeV.  
We stress that there has been no adjustment made in any parameter to bring
the earlier determined `experimental' points shown in figure \ref{EOSSfix}
into the remarkable agreement with the properties of equations of state of
the quark-gluon plasma, also seen  
in figure \ref{Ebfix}\,.

\begin{figure}[tb]
\vspace*{0.cm}
\hspace*{-1.cm}\psfig{width=14.2cm,clip=,figure=\pathnow 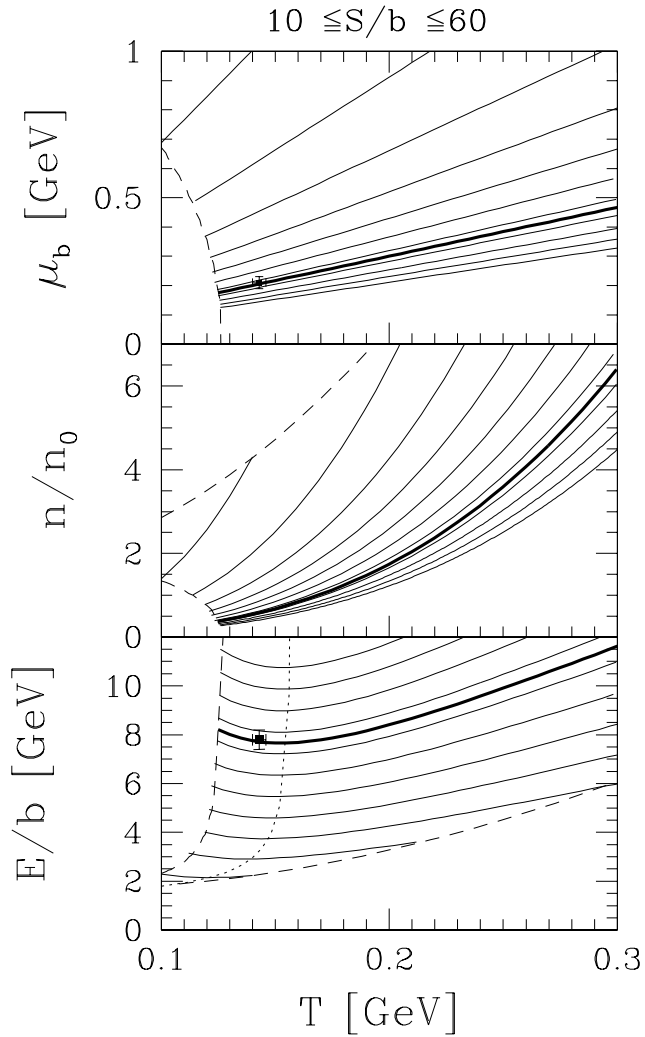}
\vspace*{-0.6cm}
\caption{ 
 From top to bottom: $\mu_b,\ n/n_0$ and $E/b$; 
lines shown correspond to fixed entropy per baryon 
$S/b= 10$ to 60 by step of 5 (left to right). 
Thick solid lines: result for $S/b=42.5$.
Limits:  energy density $\varepsilon_{\mathrm q,g}=0.5$\,GeV/fm$^3$ and
baryo-chemical potential $\mu_b=1$\,GeV. The experimental points denote
chemical freeze-out analysis result \protect\cite{Let00}, 
discussed in section \protect\ref{analyze}.
\label{EOSSfix}
}
\end{figure}
 
The trajectories at fixed energy per baryon $E/b$ are shown in the
$T$--$\lambda_q$ plane in figure \ref{Ebfix}, for the values 
(beginning at right) $E/b=2.5$ to 9.5\,GeV by step of 1. 
The highlighted curve 
corresponds to the value $E/b=7.8$\,GeV which is the local intrinsic 
energy content of the hadronizing QGP fireball formed at SPS
Pb--Pb interactions at the projectile energy 158$A$ GeV \cite{Let00}.
Dotted line, in figure \ref{Ebfix}, corresponds to $P=0$, the 
solid line that follows it is the phase transition
line where the QGP pressure is balanced by pressure of the hadron gas. 
This is the  condition at which the equilibrium transition would occur 
in a slowly evolving system, such as would be the early universe. 
The properties of hadronic gas are obtained resuming numerically 
the contribution of all known hadronic particles including resonances, 
which effectively accounts for the presence of interactions in the
confined hadron gas phase \cite{HAG}.

\begin{figure}[tb]
\vspace*{-3.8cm}
\hspace*{-1cm}\psfig{width=10.5cm,clip=,figure=\pathnow 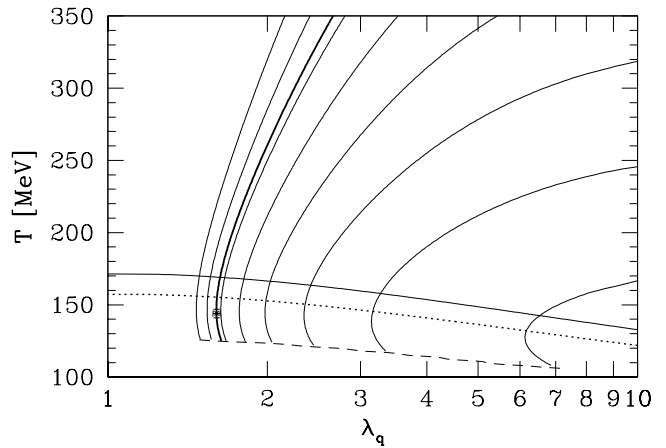}
\vspace*{-0.6cm}
\caption{ 
Contours of constant energy per baryon in QGP 
in the $T$--$\lambda_q$ plane: From right to 
left $E/b=2.5$ to 9.5\,GeV by step of 1, thick solid line is for 
$E/b=7.8$\,GeV.
Dotted line corresponds to $P=0$, above this, the solid line is the 
phase transition where the QGP pressure is balanced by pressure 
of the hadron gas. The experimental point denotes 
chemical freeze-out analysis result \protect\cite{Let00},
see section \protect\ref{analyze}. Bottom dashed line
boundary is at energy density 0.5\,GeV/fm$^3$.
\label{Ebfix}
}
\end{figure}

In relativistic nuclear collisions, the formation of equilibrium state competes
with the evolution of the fireball. The slowest of all the equilibria is
certainly the chemical equilibration of strange quarks \cite{Raf82}. 
Next slowest is the equilibration of light quarks. The chemical relaxation 
time constant for the production of light quarks has been obtained 
in the first consideration of chemical QGP equilibration, see 
figure 2 in \cite{Raf82}: $\tau_{GG\to q\bar q}(T=250\,\mbox{MeV})=0.3$\,fm.
The chemical equilibration of gluons has also been questioned, and is 
found to be slow, when only $2G\to 3G$ processes are allowed \cite{Bir93}. 
 Since gluon  fusion processes are proportional to the square of 
gluon abundance, $\tau_{GG\to q\bar q}(T=250\,\mbox{MeV})\to 1.2$\,fm 
for processes occurring when  gluons are at 50\% of equilibrium abundance. 
This  explains why quarks trail gluons in the approach 
to equilibrium, which are struggling to equilibrate by multi-gluon
production processes \cite{Shu92}. It is generally assumed that the
approximate thermal (kinetic) equilibrium is established
much faster. The mechanisms of this process remain under 
investigation \cite{Bia99}.

In  figure \ref{TLinit}, a  case study how the chemical equilibration of 
quarks cools the gluon-chemically equilibrated  fireball is presented.
The change of both the initial temperature $T_i$ (upper panel) and $\lambda_q$
(lower panel) as function of $n_f$, for $E/b=9.3$\,GeV, the total final 
energy content of Pb--Pb collisions
at CERN, is presented.  We see that in the initial state an equilibrated glue 
phase at $T\simeq270$\,MeV may have been reached. Full chemical equilibrium 
would correspond to $T\ge 230$\,MeV, but more likely this value is 
somewhat reduced due to flow dilution that has occurred in the process 
of chemical equilibration.

\begin{figure}[tb]
\vspace*{-3.7cm}
\hspace*{-0.6cm}\psfig{width=12cm,clip=,figure=\pathnow 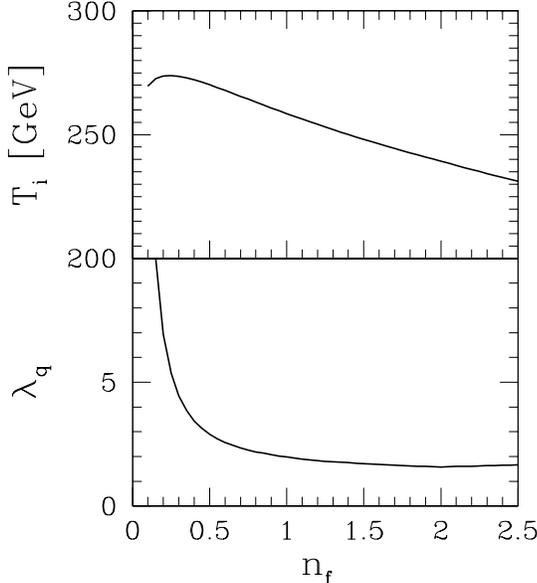}
\vspace*{-1.2cm}
\caption{ 
The initial conditions as function of the number of Fermi degrees of freedom
$n_f$. Upper panel for temperature and lower for quark 
fugacity, at $E/b=9.3$\,GeV.
\label{TLinit}
}
\end{figure}

One of the interesting quantities is the QGP energy density 
which we show in figure \ref{ESBFix}, for both fixed $E/b$ (top) and 
fixed $S/b$ bottom, for fully equilibrated condition $n_q=2, n_s=0.5$. 
We see that for $E/b>6$\,GeV and $S/b>25$ the influence of 
finite baryo-chemical potential is minimal and the lines coalesce. 
In other words, at conditions we encounter at SPS, we can correlate 
the energy density with  temperature alone $\varepsilon \simeq aT^4$, as
seen for $n_f=3>2.5$ in figure \ref{figendens25}.   
In figure \ref{ESBFix}, we see that the energy density 3\,GeV/fm$^3$ is 
established when the temperature in the equilibrated fireball equals 212\,MeV. 
Considering the observed high inverse slopes of strange particles, 
one can assume that the 
the plasma phase, before it reached full chemical equilibrium, has been
 at about $T\simeq 250$\,MeV and the energy density in this state has
most likely been still higher, above 4\,GeV/fm$^3$. We believe that 
our evaluation of the properties of the QGP liquid is, at this range
of temperature $T\simeq 1.5 T_c$, reliable. 

As a final step in the study of the properties of the QGP liquid, 
 we consider the conditions 
relevant for the formation of the QGP, and consider the behavior
for $n_f=1$. We show, in figure \ref{Ebnf1fix}, lines
of fixed energy per baryon $E/b=$3,\,4,\,5,\,6,\,8,\,10,\,20,\,50 and 100\,GeV,
akin to results we have shown for $n_f=2+0.5$, in figure \ref{Ebfix}. The 
horizontal solid line is where the equilibrated hadronic gas phase has the same 
pressure as QGP-liquid with  semi-equilibrated  quark abundance. The 
free energy of the QGP liquid must be lower (pressure higher) 
in order for  hadrons to dissolve into the plasma phase. The dotted lines  
in figure \ref{Ebnf1fix}, from bottom to 
top, show where the pressure of the semi-equilibrated QGP phase is equal to 
$\eta=$ 20\%,\,40\%,\,60\%,\,80\% and 100\%,\, $\eta$
being the `stopping' fraction of the dynamical collisional pressure \cite{acta96}:
$$P_{\mathrm{col}}=\eta\rho_0\frac{P_{\mathrm{CM}}^2}{E_{\mathrm{CM}}}\,.$$

\begin{figure}[tb]
\vspace*{-1.cm}
\hspace*{-1.3cm}\psfig{width=14.5cm,clip=,figure=\pathnow 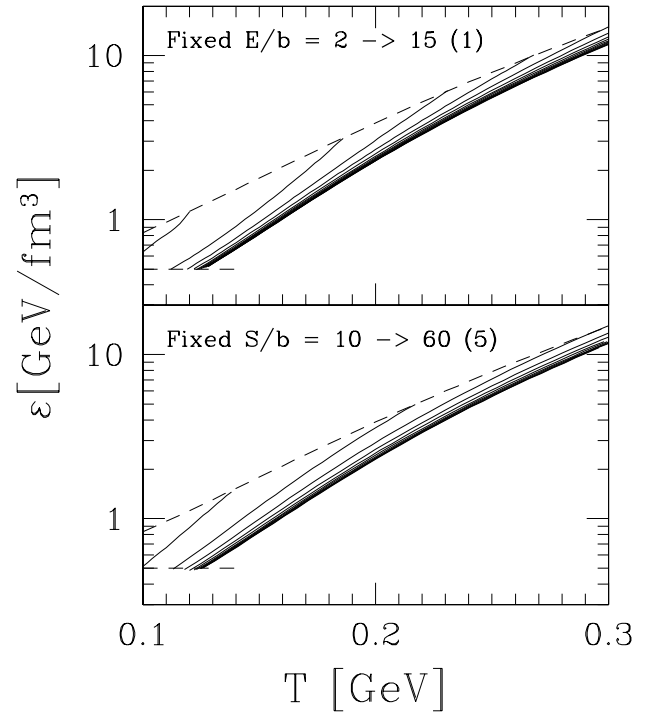}
\vspace*{-4.2cm}
\caption{ 
Energy density in  QGP as function of temperature. Top panel:
for a fixed Energy per baryon $E/b$ = 2 to 15\,GeV by step of 1; 
bottom panel for fixed entropy per baryon $S/b=$ 10 to 60 by step of 5. Boundaries are
$\varepsilon=0.5$\,GeV/fm$^3$ at the bottom and $\mu_b=1$\,GeV at the top.
\label{ESBFix}
}
\end{figure}

The rationale to study, in figure \ref{Ebnf1fix}, lines at fixed $E/b$ is that, 
during the nuclear collision which lasts about 
$2R_N/\gamma_L2c\simeq 13/18$\,fm/$c$, where $\gamma_L$ is the Lorentz factor between 
the lab and CM frame and $R_N$ is the nuclear radius, parton collisions 
lead to a partial (assumed here to be 1/2) chemical equilibration of the hadron 
matter. At that time,  the pressure exercised corresponds to collisional pressure
$P_{\mathrm{col}}$\,. This stopping fraction, 
seen in the transverse energy produced, is about 40\% for S--S collisions 
at 200$A$ GeV and 60\% for Pb--Pb collisions at 158$A$ GeV. If the momentum-energy
and baryon number stopping are similar, as we see in the experimental 
data, then the SPS collisions at 160--200$A$ GeV are found in the highlighted
area left of center of the figure. In the middle of upper boundary of this area, we
would expect the beginning evolution of the thermal but not yet 
chemically equilibrated Pb--Pb fireball, and in the lower
left corner of the S--S fireball. We note that the temperature reached in S--S 
case is seen to be about 25 MeV lower than in the Pb--Pb case.
The lowest dotted line (20\% stopping) nearly coincides with the non-equilibrium phase
boundary (solid horizontal line, in figure \ref{Ebnf1fix}) 
and thus we conclude  that this is, for the condition $n_f=1$, the lowest
stopping that can lead to formation of a deconfined QGP phase. Such a low stopping would
be encountered possibly in lighter than S--S collision systems or/and at large impact 
parameter interactions of larger nuclei.

The highlighted area, right
of center of the figure \ref{Ebnf1fix}, 
corresponds to the expected conditions in Pb--Pb 
collisions at 40$A$ GeV. If we assume that the stopping here is near 80\%,
then the initial conditions for fireball evolution would be found towards the 
upper right corner of this highlighted area. We recognize that the
higher stopping nearly completely compensates the effect of reduced 
available energy in the collision and indeed, we expect that we form QGP 
also at these collision energies. It is important to realize that
we are entering a domain of parameters, in particular $\lambda_q$, 
for which the extrapolation of the  lattice results is not 
necessarily reliable, and thus our equations of state have some systematic
uncertainty.   

\begin{figure}[tb]
\vspace*{-3.6cm}
\hspace*{-1cm}\psfig{width=10.5cm,clip=,figure=\pathnow 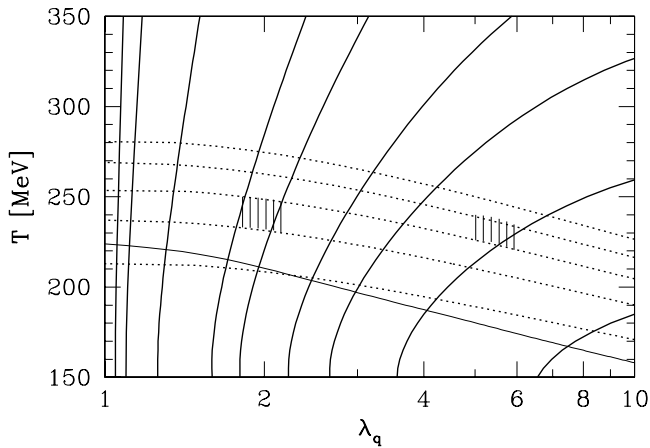}
\vspace*{-0.6cm}
\caption{ 
Contours of energy per baryon in QGP in the $T$--$\lambda_q$ plane 
for $n_f=1$: From right to left 
$E/b$=3, 4, 5, 6, 8, 10, 20, 50 and 100\,GeV\,.
Thin, nearly horizontal line: hadronic gas phase has the same pressure 
as the QGP-liquid with semi-equilibrated quark flavor. Dotted lines from
bottom to top: pressure in QGP liquid equals 20\%,\,40\%,\,60\%,\,80\%,\, 
and 100\%, of the dynamical collisional pressure.
\label{Ebnf1fix}
}
\end{figure}

\section{Experimental data analysis and QGP-EoS}\label{analyze}
A full account of our prior analysis of the 158$A$ GeV Pb--Pb collision
system has appeared \cite{Let00,Let99}. We briefly summarize the results
that we require for the study of QGP properties at fireball breakup. 
In table \ref{fitqpbs}, in upper section, we present the parameters
$T_f$ (the chemical freeze-out temperature), $v_c$ (the collective flow 
velocity at sudden breakup), $\lambda_{q}$ (the light quark fugacity),
$\lambda_{s}$  (the strange quark fugacity), $\gamma_{q}$ (the light
quark phase space occupancy), $\gamma_{s}$ (the strange quark
phase space occupancy). These are derived from analysis of all hadrons 
excluding $\Omega$ and $\overline\Omega$, which data 
points are not following the same systematic production pattern. 
These parameters characterize completely the physical properties of the 
produced hadrons, and these properties
are shown in the bottom section of table \ref{fitqpbs}.

In the heading of the table, the
total error, $\chi^2$ is shown, along with the number of data points $N$, parameters 
$p$ and data point constraints $r$. The confidence level that is
reached in our description is near or above 90\%, depending on 
scenario considered.  
The scenarios we consider are seen in the
columns of table \ref{fitqpbs}: an unconstrained
description of all data in the first column, constraint to exact strangeness
conservation in the observed hadrons, second column. Since in both
cases the parameter $\gamma_{q}$ assumes value that maximizes the 
entropy and energy content in the pion gas, we assume this value 
in the so constrained third column.

We can now  check the consistency between the statistical parameters 
(top panel of table \ref{fitqpbs}) and the 
physical properties of the fireball (bottom  panel of table \ref{fitqpbs})
which are maintained in the process of 
hadronization.  We note that the energy shown in this table, is the intrinsic 
energy in the flowing frame. The CM-laboratory energy includes the kinetic 
energy of the flow and thus is greater, to be  obtained 
multiplying the result shown in table \ref{fitqpbs} 
by the Lorentz factor $\gamma=1/\sqrt{1-v^2_c}=1.19$.
Thus the initial value of the energy per baryon that the 
system has had before expansion started has been $E^0/b\simeq 9.3$\,GeV,
as used in figure \ref{TLinit} and in the estimate presented in figure \ref{Ebnf1fix}.

\begin{table}[htb]
\caption{\label{fitqpbs}
Results of study of Pb--Pb hadron production \protect\cite{Let00}: 
in the heading: the total quadratic relative error 
$\chi^2_{\rm T}$, number of data points $N$, parameters $p$ and 
redundancies $r$;  in the upper section: statistical model parameters
which best describe the experimental results for Pb--Pb data.
Bottom section: specific energy, entropy, anti-strangeness, net strangeness
 of  the full hadron phase space characterized by these
statistical parameters. In column one, all statistical parameters and 
the flow vary. In column two, we fix $\lambda_{s}$ by requirement of 
strangeness conservation, and in column three, we fix $\gamma_{q}$ at
the pion condensation point $\gamma_{q}=\gamma_{q}^c$.}

\begin{center}
\begin{tabular}{l|ccc}
                       & Pb$|_v$            & Pb$|_v^{\rm sb}$ & Pb$|_v^{\rm sc}$     \\
$\chi^2_{\rm T};\ N;p;r$&2.5;\ 12;\,6;\,2   & 3.2;\ 12;\,5;\,2 & 2.6;\ 12;\,5;\,2     \\
\hline
$T_{f}$ [MeV]          &    142 $\pm$ 3     &  144 $\pm$ 2     &  142 $\pm$ 2       \\
$v_c$                  &   0.54 $\pm$ 0.04  & 0.54 $\pm$ 0.025 & 0.54 $\pm$ 0.025   \\
$\lambda_{q}$          &   1.61 $\pm$ 0.02  & 1.605 $\pm$ 0.025& 1.615 $\pm$ 0.025   \\
$\lambda_{s}$          &   1.09 $\pm$ 0.02  & 1.10$^*$         & 1.09 $\pm$ 0.02      \\
$\gamma_{q}$           &   1.7 $\pm$ 0.5    & 1.8$\pm$ 0.2   &${\gamma_{q}^c}^*=e^{m_\pi/2T_f}$ \\
$\gamma_{s}/\gamma_{q}$&   0.79 $\pm$ 0.05  & 0.80 $\pm$ 0.05  & 0.79 $\pm$ 0.05     \\
\hline
$E_{f}/B$              &   7.8 $\pm$ 0.5    & 7.7 $\pm$ 0.5    & 7.8 $\pm$ 0.5     \\
$S_{f}/B$              &    42 $\pm$ 3      & 41 $\pm$ 3       & 43 $\pm$ 3        \\
${s}_{f}/B$            &  0.69 $\pm$ 0.04   & 0.67 $\pm$ 0.05  & 0.70 $\pm$ 0.05    \\
$({\bar s}_f-s_f)/B\ \ $   &  0.03 $\pm$ 0.04   & 0$^*$            &  0.04 $\pm$ 0.05   \\
\end{tabular}
\end{center}
\end{table}

In the bottom panel in figure~\ref{EOSSfix}, we saw that the Temperature 
$T_f=143\pm3$\,MeV and  energy per baryon $E/b=7.8$\,GeV 
where just at $S/b=42.5$ seen  table \ref{fitqpbs}. Similarly, in the top 
panel, the baryo-chemical potential $\mu_b=3 T_f\ln \lambda_q=204\pm10$\,MeV  is 
 as required for the consistency of QGP properties. 
A similarly embarrassing agreement of the hadron yield analysis results with 
properties of the QGP fireball is seen in    figure~\ref{Ebfix}, but
that is just a different representation (at fixed $E/b$) of the result we saw at 
fixed $S/b$, in figure~\ref{EOSSfix}. However, the importance of this result is 
that the plasma breakup point appears  
well below  the phase transition temperature line (thin horizontal solid line). 
As this discussion shows, the properties of the QGP liquid at the 
point of hadronization  are the same as found
studying the properties of hadrons emerging from the exploding fireball. 
Both the specific energy and entropy
content of the fireball are consistent with the statistical parameters 
$T_f$ and $\mu_b$ according to our equations of state of the quark-gluon 
liquid. The freeze-out point at 
$T\simeq 143$\,MeV, seen in bottom panel of figure \ref{EOSSfix},
corresponds to an energy density $\varepsilon_f\simeq 0.6$\,GeV/fm$^3$. 
This is the value for the super-cooled plasma, the 
equilibrium phase transition occurs at twice this value, 
$\varepsilon_p\simeq 1.3$\,GeV/fm$^3$. 

We can also evaluate the hadronic phase space
energy density. We need to introduce the  excluded volume correction  \cite{HR80}. 
Considering that the point particle 
phase space energy density $\varepsilon_{pt}= 1.1$\,GeV/fm$^3$, we obtain
$\varepsilon_{HG}\simeq 0.4$\,GeV/fm$^3$, using the  value of 
${\cal B}=0.19$\,GeV/fm$^3$. Taking into account the numerous uncertainties in 
the understanding of the excluded volume effect, we conclude that the 
fireball energy density is comparable in magnitude  to the energy density
present in hadronic phase space.  We also note that the 
equilibrium phase transition curve we had presented in
figure \ref{Ebfix} was computed without
the excluded volume effect. When allowing for this, 
the solid line (critical curve) moves about half way towards 
 the dotted line where $P_{\mbox{\scriptsize QGP}}=0$.

For a vanishing baryo-chemical potential, $\lambda_q=1$, we determine
in figure~\ref{Ebfix}, that the 
phase transition temperature for point hadrons is  $T_p\simeq 172$\,MeV. 
The super-cooled  
$P=0$ temperature is at $T_c=157.5$\,MeV (dotted line at 
$\lambda_q=1$, in figure~\ref{Ebfix}), and 
an expanding fireball can super-cool to as low as $T\simeq 145$\,MeV, 
where the mechanical instability occurs \cite{Cso94,Raf00}.
We also have seen this result in the bottom panel of figure~\ref{EOSSfix} 
where the doted line, corresponding to pressure zero,  is to the right of 
the `experimental' point.

The collective velocity $v_c$ of the exploding matter remains large
in deeply super-cooled 
conditions even though the expansion slows down 
and kinetic energy is transfered back from flow to  thermal 
component, once the dotted $P=0$ line is crossed, as can be seen in 
figure \ref{EOSSfix}.  Interestingly, our central 
value $v_c=0.54$  is the  velocity of sound of the exploding fireball:
\begin{equation}
\label{vsound}
v_s^2=\left.\frac{\partial P}{\partial \varepsilon}\right\vert_{S/b}\,.
\end{equation}
The theoretical line shown in figure \ref{PLVST42} is computed for
 $S/b=42$.

\begin{figure}[tb]
\vspace*{-1.8cm}
\hspace*{-1.1cm}\psfig{width=9.5cm,clip=,figure=\pathnow 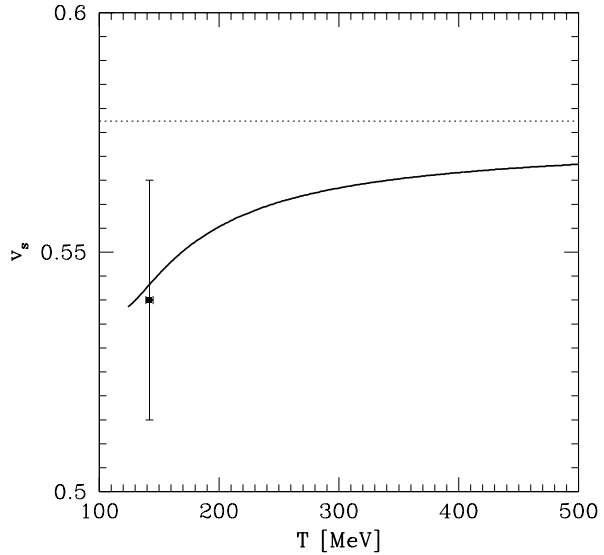}
\vspace*{-0.5cm}
\caption{ 
The velocity of sound of quark-gluon liquid at $S/b=42$\,.
 The dotted line corresponds to the value of the sound velocity of an
 ideal relativistic gas, $v_s=1/\sqrt{3}$.
\label{PLVST42}
}
\end{figure}

There is yet more evidence for the QGP nature of the fireball. It has been 
argued before that  the values of the three other
chemical parameters $ \lambda_{s}, \gamma_{q}$ and $\gamma_{s}$ suggest 
as source a deconfined QGP  fireball source \cite{Let00}:\\
{\bf  a)} The value of strange quark 
fugacity $\lambda_{s}$ can be obtained  
from the  requirement that strangeness  balances,
$\langle N_{s}-N_{\bar s}\rangle=0\,,$
which for a source in which all $s$ an $\bar s$ quarks are unbound and thus 
have symmetric phase space, implies $\lambda_{s}=1$\,.
 However, the Coulomb distortion
of the strange quark phase space plays an important role in the
understanding of this constraint for Pb--Pb collisions, 
leading to the Coulomb-deformed value $\lambda_{s}\simeq 1.1$, which is identical 
to the value obtained from experimental data analysis $\lambda_s^a=1.09\pm0.02$\,.\\
{\bf  b)} The  phase space occupancy of light quarks 
$\gamma_{q}$ is, before gluon fragmentation, near or at 
the equilibrium value $\gamma_{q}=1$. However, as measured
by hadron abundances it is expected to  significantly exceed unity 
to accommodate the contribution from gluon fragmentation into light quark 
pairs.  There is an upper limit:
$\gamma_{q}<\gamma_{q}^c\equiv e^{m_\pi/2T_f}\simeq 1.67$\,,
which arises to maximize the entropy density in the confined hadron phase. \\
{\bf  c)} The strange quark phase space occupancy 
$\gamma_{s}$ can be computed within the framework 
of kinetic theory and is mainly influenced by  strangeness pair production 
in gluon fusion\cite{Raf82}, in early stages of the collision at high temperature, and
by dilution effect in which the already produced strangeness over saturates the
`thiner' low temperature phase space. Moreover, some gluon fragmentation also
enriches $\gamma_s$ as measured by hadron abundance. We note that some
earlier studies implicitly address the parameter $\gamma_s/\gamma_q$ which therefore
is stated in  table \ref{fitqpbs}. 

It is  worthwhile to recall here that the 
strangeness yield,  $s/b=\bar s/b\simeq 0.7$, predicted early on as the 
result of QGP formation \cite{Raf82} is also one of the
results of  data analysis seen in table \ref{fitqpbs}. This result is
found with modern kinetic studies of strangeness
production in QGP \cite{Let00,Let99,acta96}.

\section{Discussion and Conclusions}\label{conclusion}
We have presented properties of the quark-gluon
plasma equations of state, and have made several applications 
pertinent to the physics of relativistic heavy ion collisions.
These included a study of the initial attainable conditions 
reached in Pb--Pb collisions at $158A$\,GeV projectile energy,
as well as an evaluation of the hypothesis that  direct hadron
emission occurs from a disintegrating QGP fireball.

We have based our description of the QCD matter
on a form obtained in perturbative expansion of thermal 
QCD, wherein we have introduced a nonperturbative 
resummation of the thermal interaction
strength. We have also introduced the non-perturbative
vacuum properties. The two parameters of our model are the vacuum 
pressure $\cal B$ and the scale $\mu(T)$ of the QCD coupling
strength. These  were chosen to 
reproduce  the latest lattice results for pressure and energy
density obtained at zero baryon density. For study of 
finite baryon density  we rely on the perturbative behavior of the 
interacting quark-gluon gas.
This is developed within the same computational approach in which the 
pressure and energy density are obtained correctly. Fortuitously, it turns 
out that the QCD interaction effects are smallest for the baryon density,
and thus, a priori, are most reliably described. Even so, the  magnitude 
of these effects increases to the level of 45\% at the phase boundary.  

A serious limitation of our approach 
is that we rely completely  on the current 
lattice results. There are open questions about the precision of 
lattice  results we model. The requirement for
continuum extrapolation \cite{Kar00}, and the use of
the relatively large $m/T=0.4$\, make our results uncertain 
at the level of 10\%. Moreover, in the deeply supercooled 
region we do not have lattice results available and the 
behavior of QGP we consider arises solely from the analytical 
properties of our QGP model. 

In the region of 
statistical parameters of interest in the discussion of the SPS 
experimental data, the detail how the baryo-chemical potential 
enters the interaction scale $\mu$ of the strong interaction 
 does not matter since  for
 $(\pi T_c)^2\simeq 50\mu_q^2$\, the $\mu_b$-contribution to the 
interaction scale is negligible compared to the $T$-contribution. 
This, however, means that our successful description of the 
SPS-experimental results does not imply that our model of QCD 
matter can safely be used in the study of properties of 
quark star matter, or the fireball matter made at AGS energies, 
where $T_c\simeq \mu_q$ \cite{RD94}.

The validity of our equation of state model is better established in
the temperature range pertinent to study of 
the initial state of the QGP fireball formed at SPS. We have found that 
during the pre-chemical equilibrium stage of light quarks, the 
so called hot glue scenario \cite{Shu92}, the QGP plasma at SPS 
has been formed at about $T\simeq 250$--$270$\,MeV,  
see figures~\ref{TLinit} and \ref{Ebnf1fix}.
By the time light quarks have also chemically equilibrated, 
the temperature  decreases to just above
$T\simeq 210$\,MeV, and the energy density in such a QGP fireball is at 
3\,GeV/fm$^3$, as can be seen in the lower panel of figure \ref{ESBFix}.

We have studied the requirement for the minimal 
collision energy, which permits the 
formation of deconfined fireball as function 
of  `stopping' in the collision. 
We have shown that a significant change in the reaction mechanism 
of colliding nuclei can be expected when the opacity is at or below 20\%,
see figure~\ref{Ebnf1fix}. At that point, formation of a deconfined phase 
fireball is not  assured as is seen comparing in the $T$--$\lambda_q$
plane the boundary between hadron phase
(solid horizontal  line in figure~\ref{Ebnf1fix}) and 
chemically non-equilibrated QGP initial conditions (lowest dashed horizontal 
line in figure~\ref{Ebnf1fix}).  Such a low stopping is expected   
for sufficiently small collision systems, with participant mass below 
that seen in central S--S interactions.
 The results shown in  figure~\ref{Ebnf1fix} imply that 
the QGP-liquid phase should be formed for Pb--Pb collisions at 
all collision energies accessible to SPS.   In fact, were it 
not for the uncertainties inherent in the extrapolation 
of lattice results  to high baryon density
 reached at AGS, we could argue that
deconfinement also occurs for Au--Au 
collisions in the high AGS energy range \cite{LTR94}. 

We have also compared the properties of the QGP phase to conditions 
present at time of hadronization, as 
obtained within an analysis of hadron production \cite{Let00}. 
This analysis determines in a first step  a set of statistical 
parameters which  describe well experimental hadron multiplicity 
results.  Since this set of parameters 
also  characterizes the phase space of all hadronic particles, 
in a second step one can estimate the energy, 
baryon number and entropy content
contained in all hadrons produced. A comparison to the QGP 
equations of state can be made, assuming that in the hadronization
of the deconfined matter there has been no reheating, and  no shift in 
chemical equilibrium properties.  This is
the  sudden hadronization. In this case we can compare 
final state fireball properties with the behavior of
QGP matter evaluated at same values of $T,\lambda_q,\lambda_s$.

However, the momentum distribution of final state hadrons could be subject 
to modification after QGP fireball breakup, which is usually expressed 
by introducing two sets of statistical parameters, the chemical
freeze-out (particle abundance freeze-out) and thermal freeze-out (spectral
shape) temperatures and chemical potentials. In our comparison of QGP
properties with fireball properties we rely on results obtained from 
an analysis of  particle multiplicity ratios, which reduce the impact
of post-hadronization system evolution. 

At the time of hadronization of a QGP fireball  gluons contribute to form 
a chemical nonequilibrium excess of hadrons. Thus even if 
the hadronizing QGP is (near) chemical
equilibrium, the confined phase in general will require allowance for chemical
nonequilibrium. A non-equilibrium analysis  is  more general
than the equilibrium models \cite{Bir00}, and always 
describes the hadron production data  better \cite{Let98}. This
result in itself constitutes evidence for the presence of primordial 
deconfined phase. Here, we have for the first time demonstrated 
that there is a good agreement between 
the energy, entropy and baryon number content of the collision 
fireball and the properties expected from the study of supercooled 
QGP phase.

Although this comparison has produced a very good agreement 
between the fireball properties and the QGP properties, 
this could be just a coincidence. Firstly, 
we do not fully understand the limitations in the description of the 
fireball matter.  The hadron  phase space is used with statistical 
populations allowing for presence of many very heavy  
hadron resonances. This hypothesis has not been experimentally 
verified at this time. In fact it is to be expected
that a more complex pattern of formation of hadronic particles arises,
and we estimate the systematic error in the computed physical 
properties of hadron phase space to be at least 15\%, arising 
from the assumed populations of unobserved hadronic states. 
Because of the nature of the QGP equations of state
the `measurement' point in figure \ref{EOSSfix} presented
at the values of $\mu_b$ and $T$ found from hadron multiplicity analysis 
would be compatible with a range  $35<S/b<50$. While it is nice to see
the evaluated hadron abundance property to be just 
it in the middle of this QGP range, we must keep in mind that a wide range 
of entropy values is  permissible. 

It must be clearly said that the fireball properties we study are 
incompatible with properties of chemically and thermally 
equilibrated confined hadron gas. Perhaps the simplest way to see this
is to realize that applying such a model to describe  strange
hadrons gives a set of parameters which fail to describe  
the total hadron yield  by many (10-15) standard deviations.
The key point of our study is  that we have shown that 
a natural way to explain 
consistently all hadronic production data obtained in $158A$\,GeV
Pb--Pb collisions is to invoke as reaction picture the formation, and 
sequel sudden hadronization,  of a QGP fireball.

\subsection*{Acknowledgments}
Work supported in part by a grant from the U.S. Department of
Energy,  DE-FG03-95ER40937\,. Laboratoire de Physique Th\'eorique 
et Hautes Energies, LPTHE, at  University Paris 6 and 7 is supported 
by CNRS as Unit\'e Mixte de Recherche, UMR7589.




\end{narrowtext}

\end{document}